\documentclass[utf8]{FrontiersinHarvard} % for articles in journals using the Harvard Referencing Style (Author-Date), for Frontiers Reference Styles by Journal: https://zendesk.frontiersin.org/hc/en-us/articles/360017860337-Frontiers-Reference-Styles-by-Journal
\usepackage{url,hyperref,lineno,microtype,subcaption}
\usepackage[onehalfspacing]{setspace}
\usepackage{gensymb}
\usepackage{mhchem}
\usepackage{amsmath}
\usepackage[flushleft]{threeparttable}

   \newcommand{\arepo}{{\sc arepo}}          % code names
   \newcommand{\athena}{{\sc athena}}          % code names
   \newcommand{\athenapp}{{\sc athena++}}          % code names
   \newcommand{\dispatch}{{\sc dispatch}}          % code names
   \newcommand{\dragon}{{\sc dragon}}          % code names
   \newcommand{\enzo}{{\sc enzo}}          % code names
   \newcommand{\flash}{{\sc flash}}          % code names
   \newcommand{\gizmo}{{\sc gizmo}}          % code names
   \newcommand{\orion}{{\sc orion}}          % code names
   \newcommand{\phantomcode}{{\sc phantom}}          % code names
   \newcommand{\pluto}{{\sc pluto}}          % code names
   \newcommand{\ramses}{{\sc ramses}}          % code names
   \newcommand{\sphNG}{{\sc sphNG}}          % code names
   \newcommand{\sfumato}{{\sc sfumato}}          % code names
   \newcommand{\zeus}{{\sc zeus}}          % code names

\newcommand{\Fig}[1]{Fig.~\ref{fig:#1}}    % Fig. reference
\newcommand{\edit}{}
\def\keyFont{\fontsize{8}{11}\helveticabold }
\def\firstAuthorLast{Kuffmeier} %use et al only if is more than 1 author
\def\Authors{Michael Kuffmeier\,$^{1}$}
\def\Address{$^{1}$Niels Bohr Institute, University of Copenhagen, 1350 Copenhagen, DK}
\def\corrAuthor{Michael Kuffmeier}

\def\corrEmail{kueffmeier@nbi.ku.dk}

\begin{document}
\onecolumn
\firstpage{1}

\title[Star-disk formation]{Magnetohydrodynamical modeling of star-disk formation: from isolated spherical collapse towards incorporation of external dynamics} 
% Theoretical results of the physical and chemical properties, observational characteristics and significance of the first and second hydrostatic core stages.

\author[\firstAuthorLast ]{\Authors} %This field will be automatically populated
\address{} %This field will be automatically populated
\correspondance{} %This field will be automatically populated

\extraAuth{}% If there are more than 1 corresponding author, comment this line and uncomment the next one.
%\extraAuth{corresponding Author2 \\ Laboratory X2, Institute X2, Department X2, Organization X2, Street X2, City X2 , State XX2 (only USA, Canada and Australia), Zip Code2, X2 Country X2, email2@uni2.edu}

\maketitle

\begin{abstract}

%%% Leave the Abstract empty if your article does not require one, please see the Summary Table for full details.
\section{}
{\edit
The formation of protostars and their disks has been understood as the result of the gravitational collapse phase of an accumulation of dense gas that determines the mass reservoir of the star-disk system.
Against this background, the broadly applied scenario of considering the formation of disks has been to model the collapse of a dense core assuming spherical spherical symmetry. 
Our understanding of the formation of star-disk systems is currently undergoing a reformation though. 
The picture evolves from interpreting disks as the sole outcome of the collapse of an isolated prestellar core to a more dynamic picture where disks are affected by the molecular cloud environment in which they form.
In this review, we provide a status report of the state-of-the-art of
spherical collapse models that are highly advanced in terms of the incorporated physics together with constraints from models that account for the possibility of infall onto star-disk systems in simplified test setups, as well as in multi-scale simulations that cover a dynamical range from the Giant Molecular Cloud environment down to the disk. 
Considering the observational constraints that favor a more dynamical picture of star formation, we finally discuss the challenges and prospects in linking the efforts of tackle the problem of star-disk formation in combined multi-scale, multi-physics simulations. 
}

\tiny
 \keyFont{ \section{Keywords:} Star formation, (magneto-)hydrodynamics, 
 disk formation phase, non-ideal MHD, Infall, accretion
 } %All article types: you may provide up to 8 keywords; at least 5 are mandatory.
\end{abstract}

\section{Introduction}
The by-now classical approach of modeling the formation of a star dates back to more than half a century to the pioneering work of \citep{Larson1969}, who started from the assumption of an isolated spherical core that collapses due to its own gravity. 
This assumption has become the standard approach in modeling the formation of individual stars and their disks that form as a result of conservation of angular momentum during the collapse phase.
However, in recent years it has become more and more clear that 
the the morphology of the precursors of stars, namely prestellar cores, often deviates significantly from spherical symmetry in turbulent filamentary Giant Molecular Clouds \citep{Andre+2014}.
In addition, asymmetric features (`streamers') \citep[see review by][]{Pineda+2023}, as well as strong indications for late infall \citep[e.g., SU Aur][]{Ginski+2021} challenge our traditional view on disk formation.
This review is an attempt to concisely summarize the developments made in developing state-of-the-art multi-physics models of spherical collapse, and put those in context to multi-scale models that account for the larger scale dynamics of the molecular cloud environment. 
We emphasize that the scope of the review is the formation process of disks.
These disks evolve over time and disperse. 
An overview of important effects regarding disk dispersal through binary interactions \citep[e.g.,][]{Kuruwita+2019,Offner+2023}, stellar flybys \citep[e.g.,][]{Cuello+2023,Smallwood+2023spiral} or external photoevaporation \citep[e.g.,][]{WinterHaworth2022} can be found in the respective references. 
\textrm{Note that developments that later marked important advancements for modeling disk formation were initially included in studies that focused on binary/multiple formation. 
Examples are the use of adaptive mesh refinement (AMR) \citep{Truelove+1998,Kratter+2010ApJ} or nested grids \citep{BurkertBodenheimer1993,MatsumotoHanawa2003} in spherical collapse simulations. 
The focus on multiplicity rather than disk properties is not surprising considering that disks easily form in hydrodynamical simulations.}

\textrm{This review focuses on results obtained through modeling of (proto-)star formation.  
The presented results are derived using various codes that adopt different methodology. 
Traditionally, star formation has either been modeled following a Lagrangian approach, i.e., smoothed particle hydrodynamics (SPH), or an Eulerian approach, i.e., a mesh/grid codes.
In SPH codes, the gas distribution is represented through particles and the properties of the gas is computed through averaging over the nearest neighbors of interacting particles. 
SPH codes that are used for modeling disk formation are \dragon\ \citep[][]{Goodwin+2004}, \phantomcode\ \citep[][]{Price+2018}, optimized versions of \sphNG\ \citep[][]{Benz+1990}, as well as Godunov SPH methods \citep{IwasakiInutsuka2011}.
In Eulerian codes, the gas is discretized in form of a mesh or grid consisting of cells, and the evolution of the gas is simulated by calculating the flux through the cell boundaries to update the cell quantities. 
The grid is typically assumed to be cartesian or spherical in star formation models.
Examples of grid codes used for disk modeling are \athena\ \citep[][]{Stone+2008}, \athenapp\ \citep[][]{Stone+2020}, \dispatch\ \citep[][]{Nordlund+2018}, \enzo\ \citep[][]{Bryan+2014}, \flash\ \citep[][]{Fryxell+2000}, \pluto\ \citep[][]{Mignone+2012}, \orion\ \citep[][]{Klein1999}, \ramses\ \citep[][]{Teyssier2002,FromangHennebelleTeyssier2006}, \sfumato\ \citep[][]{Matsumoto2007} and \zeus\ \citep[][]{StoneNorman1992}.
While SPH codes intrinsically adapt to resolve the higher densities during the star formation process, 
many grid codes offer the possibility to resolve the process by the use of a grid with flexible cell sizes.
The resolution of the grid can either be fixed at the beginning of the simulation in form of static or flexible through AMR. 
Static grids are often used for spherical or nested grids, where the forming star is located at the center such that by construction the gas close to the star is resolved with higher resolution than the gas at larger radial distances.
Nowadays, the traditional distinction between SPH and grid codes has become increasingly softened by the development of methods that contain properties of both approaches such as moving mesh codes \citep[\arepo][]{Springel2010Arepo,Weinberger+2020} and codes allowing for the use of meshless methods \citep[\gizmo][]{Hopkins2014}.  
Each method has its benefits and disadvantages. 
On the one hand for instance, it is by construction straight-forward to handle advection of flows with SPH algorithms, while more care is required in grid codes. 
On the other hand, the use of constrained transport in grid codes \citep{EvansHawley1988,BalsaraSpicer1999,LondrillodelZanna2004} guarantees the absence of unphysical magnetic monopoles ($\nabla \cdot \mathbf{B}=0$), whereas a careful implementation of divergence-cleaning is required to achieve this condition in SPH \citep[see for instance][]{Tricco+2012,Tricco+2016}.}

\textrm{For more details on the methods that are commonly used in star formation modeling, we refer the reader to dedicated reviews. 
For the technicalities of SPH, see for instance \cite[][in this volume]{Tricco2023} or previous reviews \citep[][]{Rosswog2009,Springel2010,Price2012}.
For an overview of the numerical methods in grid-based codes, we refer to \cite{Teyssier2015,TeyssierCommercon2019} as well as to the respective papers corresponding to the individual codes.
Alternatives to the more traditional methods are presented and discussed in \cite{Hopkins2015}.
To highlight which method was used in the references mentioned in this review, however, we mark works that used grid codes in \textcolor{MidnightBlue}{blue} color and those using SPH in \textcolor{violet}{violet} from here on. 
Studies that used none or both methods are referred to in regular font color. }

Section 2 of this review provides an overview of disk formation in spherical collapse with a focus on how to solve the magnetic braking problem. 
Section 3 focuses on the dynamics beyond the prestellar core and how modeling can be used to account for the effect of infall onto the disk formation process. 
Section 4 summarizes the results and provides an outlook on how multi-scale, multi-physics models can help to understand the small-scale subtleties of disk formation that is important for planet formation in the context of the larger-scale dynamics.

\section{Formation of disks in models of isolated spherical collapse}
In general, the assumption of spherical collapse is the obvious first choice for modeling gravitational collapse. 
The assumption of spherical symmetry simplifies the three-dimensional spatial problem to a one-dimensional problem.
Moreover, it allows to study the effect of various parameters in a well-defined setup.
Over the years, more and more parameters were added to the setup to test their effect on the collapse phase. 
An important additional parameter of the setup was to initialize the sphere with a net rotation rate. 
Theory predicts that during the collapse, the vertical velocity components cancel out each other, but the net rotational component remains such that the collapse leads to the formation of a rotationally supported disk as a consequence of angular momentum conservation. 
Purely hydrodynamical models (i.e., models without magnetic fields) demonstrated the formation of disks with smaller/larger disk sizes for initially weak/strong rotation of the sphere \textrm{using smoothed particle hydrodynamics (SPH) \textcolor{violet}{\citep{Bate1998,Bate+2010_HDRTdisk,Walch+2009,Walch+2010}}, adaptive mesh refinement (AMR) \textcolor{MidnightBlue}{\citep{Truelove+1998,Bannerjee+2004}} or nested grid codes \textcolor{MidnightBlue}{\citep{Yorke+1993,Saigo+2008,Machida+2010}}}.

To prevent very low time steps during the collapse, it is common practice in many models to introduce a sink particle that is created at the center once a critical density is exceeded and possible additional criteria are fulfilled. 
During the further evolution, the sink accretes mass from the surrounding gas according to a prescribed recipe \citep[for a detailed overview on numerical methods including sink particles, please refer to the review by][]{TeyssierCommercon2019}.

\begin{figure}[h!]
\begin{center}
\includegraphics[width=10cm]{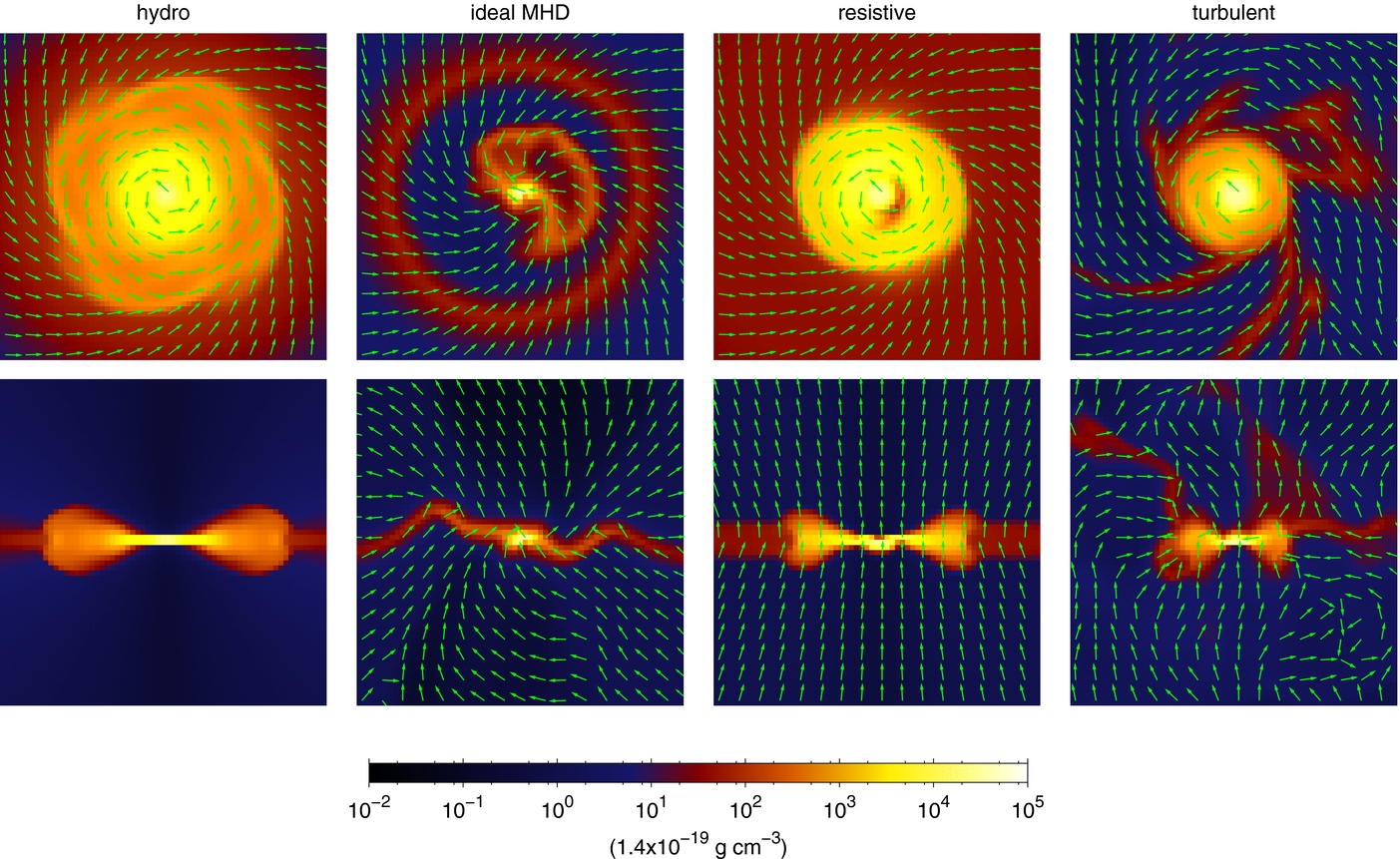}% This is a *.eps file
\end{center}
\caption{Illustration of the effect of initial conditions on the collapse phase. From left to right: hydrodynamical case (no $B$-field), ideal MHD with $B$-field, non-ideal MHD (ambipolar diffusion and ohmic dissipation), and ideal MHD with turbulence. 
\textrm{The figure is adopted from} \textcolor{MidnightBlue}{\cite[][]{Santos-Lima+2012}}.
}\label{fig:Santos-Lima}
\end{figure}

As the next step, models started to account for the presence of magnetic fields in molecular clouds \citep{Crutcher+2012} by carrying out magnetohydrodnamical simulations. 
To account for magnetic fields in the setup, the sphere is typically initialized with a magnetization defined by the ratio of enclosed mass $M$ and magnetic flux threading the sphere $\Phi=\pi R^2 B$, where $B=|\mathbf{B}|$ is the magnitude of the magnetic field strength. 
It is common practice to state the magnetization relative to the magnetized mass  
\begin{equation}
    M_{\Phi} = c_{\Phi} \frac{\Phi}{\sqrt{G}},
\end{equation}
with a numerical constant $c_{\Phi}$ and gravitational constant $G$ in terms of the normalized mass-to-flux ratio 
\begin{equation}
    \mu_{\Phi} = \frac{M}{M_{\Phi}}.
\end{equation}
According to theory, cores are magnetically supported against collapse for $\mu_{\Phi}<1$, whereas the core undergoes collapse for $\mu_{\Phi}>1$.
$c_{\Phi}$ depends on the exact configuration of the field.
In the case of a sphere threaded by a uniform parallel magnetic field, $c_{\Phi}$ is $0.126$ \citep{MouschoviasSpitzer1976}, and for the case of a field with a constant-mass-to-flux ratio, $c_{\Phi}=0.17$ \citep{Tomisaka+1988}, which is similar to $c_{\Phi}=1/(2\pi)$ for a uniformly magnetized sheet \citep{NakanoNakamura1978}.
In the latter case, the mass-to-flux ratio becomes
\begin{equation}
    \mu_{\Phi} = 2 \pi \sqrt{G} \frac{M}{\Phi},
\end{equation}
but in this section we focus on spherical collapse.
The first generations of these collapse simulations considered the ideal MHD case of sufficiently ionized gas such that the magnetic field is well coupled to the bulk neutral gas. 
The induction equation corresponding to ideal MHD is
\begin{equation}
    \frac{\partial \mathbf{B}}{\partial t} = \nabla \times (\mathbf{v} \times \mathbf{B})
\end{equation}
with bulk velocity of the gas velocity $\mathbf{v}$. 
As the magnetic field lines are perfectly coupled to the gas motion, they are dragged towards the center of mass during the collapse, which yields a characteristic hour-glass shape. 
\textrm{While the gas collapse, magnetic pinching induces the formation of a flattened structure around the forming protostar. 
In contrast to the Keplerian disk that is forming around the star in the hydrodynamical cases, the flattened structure is not rotationally supported and therefore referred to as 'pseudodisk' in the analytical \citep{GalliShu1993I} and numerical works by \textcolor{MidnightBlue}{\cite{GalliShu1993II}}. 
The pseudodisk is larger in size than the rotationally supported disk, and in the classical picture, it marks a transition zone between the protostellar environment and the rotationally supported disk \textcolor{MidnightBlue}{\citep[\textcolor{black}{e.g.,}][\textcolor{black}{for a recent study exploring the properties of pseudodisks}]{Vaisala+2023}}. 
In this review, we focus on the formation of rotationally supported disks, and for simplicity only refer to them as disks.}
When the magnetic field lines get wrapped up in azimuthal direction, they exert a torque opposing the rotation of the disk \citep{LustSchluter1955}.
In other words, magnetic fields provide a form of angular momentum transport to slow down rotation, and it is hence referred to as `magnetic braking' \citep{Mestel1968}. 
If the magnetic torque is sufficiently high, it quenches the formation of the disk.
This scenario is referred to as `magnetic braking catastrophe'.
Analytical work by \cite{Joos+2012} showed that this is the case for $\mu_{\Phi} \lesssim 10$ in good agreement with models \textrm{adopting a 2D grid} \textcolor{MidnightBlue}{\citep{Allen+2003,MellonLi2008}}, \textrm{3D grid} \textcolor{MidnightBlue}{\citep{Machida+2005,Galli+2006,HennebelleFromang2008,DuffinPudritz2009,Seifried+2011,Santos-Lima+2012}} \textrm{and 3D SPH} \textcolor{violet}{\citep{PriceBate2007}}.
While these earlier MHD simulations did not account for radiation of the forming protostar, quenching of disk formation has proven to be a robust result independently of incorporating radiative transfer schemes for $\mu_{\Phi} \lesssim 10$ \textrm{in grid} \textcolor{MidnightBlue}{\citep{Boss1997,Boss1999,Boss2002,Commercon+2010,Tomida+2010,Tomida+2013,Tomida+2015,Tomida2014,Vaytet+2018}} \textrm{or SPH codes} \textcolor{violet}{\citep{Bate+2014,Tsukamoto+2015OhmAD}}.
Exemplarily, the second column in \Fig{Santos-Lima} illustrates the magnetic braking catastrophe. 
	
Until the late 2000s, including magnetic fields into the setup was therefore considered catastrophic for disk formation,
which was at tension with the growing evidence of Keplerian disks in observations \citep{Brinch+2007,Lommen+2008,Jorgensen+2009,Lee2011,Murillo+2013,Harsono+2014,Codella+2014}. 
At the same time, magnetic fields are the most promising candidate for producing these fast jets and low-velocity disk winds \citep{Moscadelli+2022,Donati+2010,Lee+2018,Bjerkeli+2016}
though winds especially at later stages of disk evolution might also be launched by photoevaporation \citep{Alexander+2014}.
While part of the material in the disk subsequently accretes onto the protostar, a substantial amount of the mass is ejected vertically from the system in narrow protostellar jets \citep{Guszejnov+2021,Guszejnov+2022},
or in disk winds that have wider opening angles \citep{Watson+2016}. 
Constraining the role of magnetic fields is challenging because they are difficult to observe.
To get an idea of the magnetic field structure in star forming regions,
one can measure the linear polarization of non-spherical dust grains as they tend to align with the underlying magnetic field due to radiative torques \citep{Sadavoy+2019,LeGouellec+2020}. 
However, dust polarization in disks seems to be dominated by effects of self-scattering within about 100 AU from the star \citep{Kataoka+2016}, (though dichroic extinction may be responsible for parts of the polarization signal in disks, too) \citep{Lee+2021}.
That means dust polarization is a good tracer of the \textit{structure} (though not on the strength) of magnetic fields on scales of the protostellar environment beyond the disk, which allows us to at least obtain loose constraints on the role of magnetic fields in the star formation process \citep{Pattle+2023}.
As these observations inform us about the presence of magnetic fields in the star-forming regions \citep{LeGouellec+2020},
there is consensus that they are the important ingredient of magnetically driven disks \citep{Lesur+2023}.

The challenging question for modelers was therefore: `how to avoid this catastrophe during spherical collapse?'
As of today, there is consensus that magnetic braking can be sufficiently reduced to allow disk formation even for spherical collapse setups with high magnetization of $\mu_{\Phi} \lesssim 10$ if at least one of the following ingredient is considered in the models: 
\begin{itemize}
    \item non-ideal MHD, or
    \item misalignment between magnetic field orientation and rotational axis, or
    \item turbulence.
\end{itemize}
We emphasize that there are already reviews \citep[e.g.,][]{WursterLi2018,Zhao+2020,Tsukamoto+2023} in the literature available that summarize the progress made in overcoming the magnetic braking catastrophe during the last $\approx$20 years.
We refer the reader to those reviews as well as to reviews in this issue \citep[e.g.,][]{Young2023} for a more in-depth coverage of the progress made in accounting for additional physics in spherical collapse models, especially related to non-ideal MHD.
The following sections provide a more comprehensive review on those efforts and the progress made.

\subsection{Non-ideal MHD}
One way of overcoming the magnetic braking catastrophe is by accounting for resistive effects (commonly referred to as `non-ideal MHD') such that the magnetic field lines are no longer tightly coupled to the bulk motion of the gas. 
In fact, dense cores in molecular clouds are weakly ionized \citep{BerginTafalla2007} and according to models the ionization rates can reach values of $n_{\rm e}/n_{\mathrm{H}_2}\sim10^{-14}$  \citep{NakanoUmebayashi1986,UmebayashiNakano1990,Nishi+1991,Nakano2002}.  
One distinguishes between three non-ideal MHD resistivities $\eta$, ohmic dissipation $\eta_{\rm O}$, ambipolar diffusion $\eta_{\rm AD}$ and the Hall effect $\eta_{\rm H}$.
Ohmic resistivity corresponds to the regime, where all charged species (electrons, ions and charged grains) are entirely decoupled from the magnetic field.
Hall resistivity describes the regime in which the electrons are coupled to the magnetic field, but the ions and charged grains are not. In other words, Hall resistivity represents the effect of ion-electron drift.
Ambipolar diffusion is active in the regime, where all three charged components are tightly coupled to the magnetic field, and hence experience a drift force induced by the neutrals that are decoupled from the magnetic fields.
Note that most generally, resistivities are tensor quantities as each resistivity can have a directional component that differs for each dimension. 
In practice, it is a fair approximation to consider the resistivities as scalar values as the directional variations are relatively small with respect to other uncertainties in dense cores.
Furthermore, it is common practice to study protostellar collapse in models adopting the following additional assumptions.
As the mass density is typically dominated by the mass of the neutral gas, and as collisions between charged and neutral particles dominate the momentum equation, it is safe to ignore pressure and momentum from the charged species. 
Considering, in addition, long evolutionary timescales of the magnetic field and the flow of neutrals compared to those of the charged ones, one can approximate the continuity equation to be dominated by the total mass, and hence only solve this equation as the continuum equation of the gas. \citep[For more details, see for instance][]{WardleKoenigl1993,CiolekMouschovias1994,MacLow+1995,WardleNg1999}. 
Under these assumptions, the major change is in the induction equation, which can be written in its modified form for non-ideal MHD 
\begin{equation}
    \frac{\partial \mathbf{B}}{\partial t} = \nabla\times (\mathbf{v}\times\mathbf{B})
    \color{red}-\nabla\times[\eta_{\rm O} (\nabla \times \mathbf{B})]
    \color{purple}-\nabla\times \{\eta_{\rm H} [(\nabla \times \mathbf{B}) \times \frac{\mathbf{B}}{|\mathbf{B}|}] \}
    \color{blue}-\nabla\times \{\eta_{\rm AD} \frac{\mathbf{B}}{|\mathbf{B}|} \times [(\nabla \times \mathbf{B}) \times \frac{\mathbf{B}}{|\mathbf{B}|}] \}\color{black}.
\end{equation}
Moreover, the energy equation for the time evolution of internal energy $u$ becomes 
\begin{equation}
    \frac{\partial u}{\partial t} = - \frac{P}{\rho} \nabla \cdot \mathbf{v} \color{red}+ \eta_{\rm O} \frac{|\nabla \times \mathbf{B}|^2}{\rho} \color{blue}+ \eta_{\rm AD} \frac{1}{\rho} \{|\nabla \times \mathbf{B}|^2 - 
    [ (\nabla \times \mathbf{B}) \cdot \frac{\mathbf{B}}{|\mathbf{B}|}]^2\}\color{black}.
\end{equation}
If expressed in terms of total energy $E_{\rm tot}=\rho \epsilon + \frac{1}{2} \rho v^2 + \frac{1}{2}|\mathbf{B}|^2$ with specific internal energy $\epsilon$ and using total pressure $P_{\rm tot}=(\gamma -1) \rho \epsilon + \frac{1}{2}|\mathbf{B}|^2$ with adiabatic index $\gamma$, the energy equation becomes 
\begin{multline}
    \frac{\partial E_{\rm tot}}{\partial t} = -\nabla \cdot \{ (E_{\rm tot} + P_{\rm tot})\mathbf{v} - (\mathbf{v}\cdot\mathbf{B}) \mathbf{B} \\
    \color{blue}- \eta_{\rm AD} \frac{[(\nabla \times \mathbf{B}) \times \mathbf{B}] \times \mathbf{B} }{|\mathbf{B}|^2} \times \mathbf{B}
    \color{purple} -\eta_{\rm H} \frac{(\nabla \times \mathbf{B}) \times \mathbf{B} }{|\mathbf{B}|} \times \mathbf{B}
    \color{red} -\eta_{\rm O} (\nabla \times \mathbf{B}) \times \mathbf{B} \color{black}\}.
\end{multline}
Note that the Hall resistivity does not affect the internal energy equation because in contrast to ohmic dissipation and ambipolar diffusion, it is a dispersive, non-dissipative process.

\subsubsection{Ohmic resistivity}
Reflecting the complexity in accounting for the effects numerically, the first non-ideal effect included in models was ohmic dissipation, followed by ambipolar diffusion and at last the Hall effect.
In the early approaches, ohmic resistivity was set to constant values \textcolor{MidnightBlue}{\citep{Shu+2006,Krasnopolski+2010}} with the goal to test which values of $\eta_{\rm Ohm}$ would be required to allow disk formation despite the presence of strong magnetic fields.  
\textcolor{MidnightBlue}{\cite{Inutsuka+2010}, \cite{MachidaMatsumoto2011}} and \textcolor{MidnightBlue}{\cite{Machida+2011}} found the formation of large disks with ohmic resistivity only at late stages of the collapse phase when most of the envelope material had accreted onto the protostar already. 
The physical explanation for this was first laid out by \textcolor{MidnightBlue}{\cite{MellonLi2008}}.
Angular momentum transport became less efficient in these models at later stages because there was almost zero gas available at larger scales to transport the material to, making magnetic braking less efficient. 
The results are also qualitatively consistent with semi-analytic results \citep{DappBasu2010,Dapp+2012} and 3D models \textcolor{violet}{\citep{Wurster+2016}}. 
The smaller disk sizes \textrm{of $\sim 10$ au derived by the latter groups} \textrm{instead of $\sim 100$ au} \textcolor{MidnightBlue}{\citep{Inutsuka+2010,Machida+2011}} are explained by the assumption of lower resistivities and a larger sink particle by \textcolor{violet}{\cite{Wurster+2016}}. 
While it was found that disks can form with high $\eta_{\rm Ohm}$, those values can typically only be reached very close to the center of the first hydrostatic core \textrm{(see papers by} \textcolor{MidnightBlue}{\cite[][]{Marchand+2016}} \textrm{and} \textcolor{violet}{\cite{Wurster+2016}}. 
However, the values at lower densities and temperatures prevalent during protostellar disk formation are expected to be significantly smaller.  

\subsubsection{Ambipolar resistivity}
The low probability of obtaining the right conditions for circumventing strong magnetic braking solely through ohmic resistivity gave rise to a revival of seriously accounting for ambipolar diffusion as it is the dominant process at lower densities present in prestellar cores. 
The idea of considering ambipolar diffusion as an important process of individual star formation dates back to the 1970s \citep{Mouschovias1976,Mouschovias1977,Mouschovias1979}, when it was studied in the context of redistributing the magnetic field to avoid the pile-up of magnetic pressure at the center due to magnetic-flux freezing that would prevent protostellar collapse \citep[see][for a more recent linear instability analysis of magnetized sheets with ambipolar diffusion]{DasBasu2021}.
Against this background, \cite{Hennebelle+2016_ADtheory} estimated that disks form with typical disk sizes around 18 au based on analytical calculations that account for ambipolar diffusion.  

%From a modeling point of view, the role of ambipolar diffusion in the process is debated.
Early 2D models by \textcolor{MidnightBlue}{\cite{MellonLi2009}} did not find evidence for disk formation induced by including ambipolar diffusion, but their setup disallowed disk formation on radii less than $\sim 10$ au because of the sink accretion recipe.
Later studies that did not introduce a sink particle found the formation of disks enabled by ambipolar diffusion.
3D models by \textcolor{violet}{\cite{Tsukamoto+2015OhmAD}} demonstrated the increase in angular momentum for the model case with ambipolar diffusion compared to the model case without it, which resulted in the early formation of a small $\sim 1$ au disk, which is in good agreement with results obtained by \textcolor{violet}{\cite{Tomida+2015,MachidaBasu2019}}.
Similarly, conducting very detailed collapse simulations of the early collapse stage, \cite{Vaytet+2018} discovered the formation of an even smaller Keplerian disk of less than $0.1$ au in radius briefly after the first collapse phase. 
It is important to point out that these disks form at a very early stage of star formation associated with the first collapse, and that this evolutionary stage is very short-lived ($\sim 10^3$ kyr). 
The review by \cite{Young2023} presents the state-of-the art of this phase in detail.

As anticipated in previous models \textcolor{MidnightBlue}{\citep{Tomida+2013}}, a follow-up study by \textcolor{MidnightBlue}{\cite{Tomida+2017}} that introduced a sink particle, demonstrated the formation of significantly larger disks $\sim 100$ au in size due to dissipation of both the envelope and the magnetic field.
This result is in agreement with 3D models by \textcolor{MidnightBlue}{\cite{Masson+2016}}, who found the formation of relatively large disks of about 80 au for collapse simulations initialized with 50 times higher gravitational to rotational energy and a mass-to-flux ratio of $\mu=5$. 
Other studies that evolve the collapse to later times find a similar trend of larger disks at later phases \textcolor{MidnightBlue}{\citep{Marchand+2020,LeeCharnozHennebelle2021,Tu+2024}} for initial core masses $\sim1$ to $\sim10$ M$_{\odot}$, as well as studies that consider the collapse of a more massive spherical core with a mass of $\sim100$ M$_{\odot}$ \textcolor{MidnightBlue}{\citep{Rosen+2019,Mignon-Risse+2021,Commercon+2022,OlivaKuiper2023I}}. 
Comparing the results of ideal MHD with results obtained accounting for ambipolar diffusion, \textcolor{MidnightBlue}{\cite{Masson+2016}} find a sharp upper limit of B-field strength of 0.1 G even at high densities of $\rho>10^{-12}$ g cm$^{-3}$ in the $\rho$-$B$ phase space induced by ambipolar diffusion, whereas in the ideal MHD case the field strength exceeds 1 G at these higher densities.
Such a plateau value is consistent with the theoretical analysis by \cite{Hennebelle+2016_ADtheory} of a collapsing singular isothermal sphere \citep[SIS][]{Shu1977}for the assumption of a commonly assumed cosmic-ray ionization rate of $\sim10^{-17}$ s$^{-1}$, and similar plateau values around 0.1 G were also found in other studies \textrm{by} \textcolor{violet}{\cite{Tsukamoto+2017}}, \textcolor{MidnightBlue}{\cite{Hennebelle+2020}}, \textcolor{MidnightBlue}{\cite{XuKunz2021}} \textrm{and} \cite{Zier+2024}.

\subsubsection{Hall effect}
The Hall effect is the most difficult resistivity to implement and therefore the one that has been less studied although its possible influence on protostellar collapse was previously pointed out and studied (semi-)analytically \citep{WardleNg1999,Wardle2004,BraidingWardle2012}.
Today there is a decent number of papers presenting the results of spherical collapse simulations that incorporate the Hall effect in 2D \textrm{grid simulations using \zeus} \textcolor{MidnightBlue}{\citep{Krasnopolsky+2011,Li+2011,Zhao+2020Hall,Zhao+2021}}, 3D \textrm{SPH simulations using a Godunov SPH code} \textcolor{violet}{\citep{Tsukamoto+2015Hall,Tsukamoto+2017}} \textrm{as well as \phantom} \textcolor{violet}{\citep{Wurster+2016,Wurster+2018HallQunstable,Wurster+2018,Wurster+2019,Wurster+2022}}, \textrm{3D AMR simulations with \ramses} \textcolor{MidnightBlue}{\citep{Marchand+2018,Marchand+2019}}\textrm{, and 3D moving mesh simulations using \arepo} \citep{Zier+2024Hall} that incorporate Hall resistivity.
In theory, the Hall resistivity can adopt positive or negative values, but in practice it is typically negative during the protostellar collapse \textrm{as shown in SPH} \textcolor{violet}{\citep{Tsukamoto+2015Hall,Wurster+2016}} \textrm{as well as in cartesian grid simulations} \textcolor{MidnightBlue}{\citep{Marchand+2016}}.
In contrast to ohmic dissipation and ambipolar diffusion, the Hall effect depends on the direction of the magnetic field.
In the case of anti-parallel alignment of magnetic field and angular momentum, it can cause spin up of the gas and lead to the formation of a larger disk.
Contrary, in the case of parallel alignment of $\mathbf{B}$-field and initial rotation, the disk is small. 
For high values of $\eta_{\rm Hall}$ the outer envelope can even become counter-rotating with respect to the rotation of the inner disk \textcolor{MidnightBlue}{\citep{Krasnopolsky+2011,Li+2011,Zhao+2020Hall,Zhao+2021}}. 
One would therefore expect a bimodal distribution of disk sizes under conditions were the Hall-effect is dominant \textcolor{violet}{\citep{Tsukamoto+2015Hall,Wurster+2016}}, if only the two extreme scenarios of parallel or anti-parallel alignment were possible.
Reflecting the key constraints obtained from non-ideal MHD models, \cite{Lee+2021} expanded the analytical model of disk formation by \cite{Hennebelle+2016_ADtheory} to additionally cover the effects of ohmic dissipation and the Hall effect in addition to ambipolar diffusion.

\subsubsection{Misalignment between magnetic field and rotational axis}
The relative orientation of initial rotational axis and $\mathbf{B}$-field direction affects the collapse phase also in other ways than the Hall effect. 
\textrm{Early studies by \textcolor{MidnightBlue}{\cite{MatsumotoTomisaka2004}} investigated misalignment of initial magnetic field orientation with respect to the angular momentum vector of the core in \textrm{ideal} MHD simulations. 
\textrm{They find that, during core collapse, the orientation of the angular momentum vector, the magnetic field, and the pseudodisk converge to align in the central region of the cloud core due to angular momentum transport through magnetic braking.}
Moreover, }
misalignment between initial rotational axis and $\mathbf{B}$-field direction can reduce the efficiency of magnetic braking as first pointed out by \textcolor{MidnightBlue}{\cite{HennebelleCiardi2009}}. 
In fact, models showed that misalignment can enable disk formation in ideal MHD models, where disk formation is quenched when the axes are initially aligned \textcolor{MidnightBlue}{\citep{Joos+2012,Krumholz+2013,Li+2013,Gray+2018}}.
\textcolor{violet}{\cite{Tsukamoto+2018}} investigated the effects of misalignment between rotational and magnetic field axis considering the cases of perpendicular and parallel orientation. 
They find three possible mechanisms in which the misalignment can affect disk formation, namely selective accretion of material with high angular momentum in the perpendicular case and vice verse, magnetic braking during the isothermal collapse phase and magnetic braking of the disk. 
The authors point out that magnetic braking during the collapse phase would yield to the opposite effect on disk formation as reported by the aforementioned groups. 
Therefore, the consensus is that this effect is negligible for disk formation. 
In the recent review chapter, \cite{Tsukamoto+2023} emphasize that the differences in disk braking is primarily responsible for enhanced disk formation in the misalignment scenario because the magnetic field acts on longer time scales to transport angular momentum in a disk that is supported by centrifugal forces compared to the much shorter free fall time corresponding to the collapse phase.
\textrm{Analyzing results obtained in the models by \textcolor{MidnightBlue}{\cite{Li+2013}}, \cite{Vaisala+2019} and \cite{Wang+2022} emphasized the formation of time-variable features extending from the disk such as spiral features in the pseudodisk.}
In summary, the combination of models allowed to increasingly cover the parameter space of spherical collapse. 
The recently updated analytical model by \cite{Lee+2024} incorporates these constraints on misalignment between $B$-field and rotational axis, in addition to non-ideal MHD effects.

\subsubsection{Turbulence}
While it is consensus that magnetic braking quenches disk formation in ideal MHD models of spherical collapse with alignment between rotational and $\mathbf{B}$-field axis, it is also clear that these are highly idealized initial conditions. 
A mechanism of circumventing catastrophic braking in ideal MHD models of spherical collapse is by introducing turbulence.
Several studies of spherical collapse demonstrated that disks can form when initializing the spherical core with a turbulent velocity field \textcolor{MidnightBlue}{\citep{Santos-Lima+2012,Seifried+2012,Seifried+2013,Joos+2013}}. 
While in theory, there is no diffusivity in the ideal MHD limit, it is in practice a common feature that occurs at the resolution limit when solving the ideal MHD equations numerically.
\textcolor{MidnightBlue}{\cite{Santos-Lima+2012}} therefore proposed that the random motions intrinsic to turbulence induce an effective magnetic diffusivity on the smallest scales via so-called turbulent reconnection. 
Adopting a resolution of about $1$ au, this artificial resistivity therefore quenches the pile-up of magnetic fields during the collapse at values below $\sim 1$ G, which leads to a similar effect as reported for ambipolar diffusion. 
In agreement with the occurrence at the resolution limit, it is also clear that it is less efficient in simulations with higher resolution of less than 1 au \textcolor{MidnightBlue}{\citep{Joos+2013}} because higher magnetic field strengths are reached during the collapse, which implies more efficient magnetic braking. 
This means there is no numerical convergence in spherical collapse ideal MHD models down to sub-au resolution, while there is in models with ambipolar diffusion, where the $\mathbf{B}$-field strength reaches a characteristic plateau value that depends on the assumed ionization rate.
Collapse models with turbulence find that the effect of turbulence is reduced when non-ideal MHD effects are incorporated as well 
(see results found by \textcolor{MidnightBlue}{\cite{Lam+2019}} and \textcolor{violet}{\cite{WursterLewis2020}}), which is interpreted as turbulent diffusion being a secondary effect for disk formation \citep{Tsukamoto+2023}.

However, it has also been pointed out by \textcolor{MidnightBlue}{\cite{Seifried+2012}}, who considered a larger ($\approx50000$ au) and more massive core ($50$ M$_{\odot}$) with the adaptive mesh refinement code code \flash\ that turbulence leads to the formation of filaments and intrinsically induces misalignment between $\mathbf{B}$-field axis and rotational axis.
Considering the fundamental role of turbulence in the star formation process in the Giant Molecular Cloud,
\textcolor{MidnightBlue}{\cite{Seifried+2012}} therefore emphasize that turbulence is the cause for deviations from symmetry such as the relative orientation between rotational and $\mathbf{B}$-field axis, which reduces the braking efficiency and helps to circumvent the magnetic braking catastrophe. 
Results obtained from SPH simulations by \textcolor{violet}{\cite{Wurster+2019nocatastrophe}}, who carried out a parameter study of a turbulent collapsing core with $50$ M$_{\odot}$ in mass, varying mass-to-flux ratios of 3, 5, 10 and 20 and even all three non-ideal MHD effects included are in good agreement with that.
In \textcolor{violet}{\cite{Wurster+2019nocatastrophe}}, the core was initialized with a random velocity distribution such that the initial mean sonic Mach number was $\textsf{M}=6.4$.
The major conclusion of this work is that there is no magnetic braking catastrophe and disks form frequently.
In agreement with \textcolor{MidnightBlue}{\cite{Seifried+2012}}, this even holds true for the case of ideal MHD without any non-ideal MHD resistivities included. 
\textrm{Taking into account for turbulence within the core and ohmic dissipation at high densities, AMR simulations with the SFUMATO code \textcolor{MidnightBlue}{\citep{Matsumoto2007}} also showed the formation of warped disks on timescales of $\sim1000$ yr as a result of misalinged infall induced by the underlying turbulence \textcolor{MidnightBlue}{\cite{Matsumoto+2017}}.}

\subsubsection{The role of the ionization rate}
Today, it is consensus that non-ideal MHD effects can resolve the magnetic braking problem of disk formation in spherical collapse simulations.
The bigger question today is how much they affect the collapse and disk formation phase. 
Especially, early models assumed resistivities relatively crudely only roughly accounting for the dependency on density.
The individual resistivities depend on the underlying physical conditions and current state-of-the-art models account for this dependency by using pre-computed values that are assigned to the local density, magnetic field strength and temperature.
The computation of the table of resistivities is done by using chemical equilibrium models.
For instance, \cite{Marchand+2016} illustrated the differences compared to the values assumed in previous studies by \textcolor{MidnightBlue}{\cite{DuffinPudritz2009}} and \textcolor{MidnightBlue}{\cite{Machida+2007}}.
Apart from that there are some differences depending on the chemical model that was used to compute the tables.
In the table by \textcolor{MidnightBlue}{\cite{Zhao+2016}} the Hall resistivity is lower than the ambipolar resistivity for a cosmic-ray ionization rate of $10^{-17}$ s$^{-1}$ at any density, whereas the Hall resistivity exceeds ambipolar resistivity and ohmic dissipation in the range of number densities of $n=10^{6}$ cm$^{-3}$ to $n=10^{10}$ cm$^{-3}$ in the table by \cite{Marchand+2016}.

While these differences between the tables are generally more subtle, another physical effect is more crucial.
It is common practice to assume non-ideal MHD coefficients that were computed for a fixed cosmic-ray ionization rate $\zeta$. 
Typical values that are assumed for non-ideal MHD models are in the range of $\zeta \sim 10^{-18}$ s$^{-1}$ to $\zeta \sim 10^{-17}$ s$^{-1}$. 
The latter value is often referred to as the canonical value dating back to estimates by \cite{SpitzerTomasko1968}. 
While early measurements of the cosmic-ray ionization rate in dense cores \citep{Caselli+1998, Padovani+2009} are in broad agreement with these values, there has been an increasing number of observations that report significantly higher values in cores at higher densities \citep{Ceccarelli+2014,Podio+2014,Fontani+2017,Favre+2018,Cabedo+2023} together with observations of significant scatter of ionization rates between different cores \citep{IndrioloMcCall2012}. 
Such an increase towards higher densities was initially hard to explain as cosmic-rays were expected to be shielded and mirrored at the relatively densities in dense cores \citep[e.g.,][]{PadovaniGalli2013}.
It is, however, consistent with more recent modeling results that predict an enhancement of cosmic-ray ionization rates through protostellar accretion shocks \citep{Padovani+2016,GachesOffner2018}.
In fact, first maps of the cosmic-ray ionization rate of a low-mass star-forming region \citep[NGC1333][]{Pineda+2024} and of two massive clumps \citep[AG351 and AG354][]{Sabatini+2023} confirm the variation and show enhanced cosmic-ray ionization values toward some of the denser gas. 
It has therefore become more and more clear that  cosmic-ray ionization rates can vary by orders of magnitude depending on the protostellar stage as well as on the environment in which the star is forming.
 
\textcolor{violet}{\cite{Wurster+2018CR}} carried out a parameter study of the very early protostellar collapse phase with all three non-ideal MHD effects using resistivities corresponding to cosmic-ray ionization rate in the range of $\zeta \sim 10^{-30}$ s$^{-1}$ to $\zeta \sim 10^{-10}$ s$^{-1}$.
For $\zeta \gtrsim 10^{-14}$ s$^{-1}$,
the models evolve similar to the ideal MHD models, while the outcome becomes diminishingly different from pure hydrodynamical models for $\zeta \lesssim 10^{-24}$ s$^{-1}$. 
Similarly, \textcolor{MidnightBlue}{\cite{Kuffmeier+2020}} demonstrated that the assumption of a higher (smaller) cosmic ray ionization rate leads to stronger (weaker) magnetic braking, and thereby to smaller (larger) disks. 
Considering the various environments in which star-formation occurs, the effect of non-ideal MHD is expected to differ significantly between the regions \citep[see also discussion and speculation about starburst galxies or the Galactic Center by][]{WursterLi2018}.

\subsection{Dust}
\subsubsection{The role of dust on the resistivities}
The resistivities are also affected by the assumed dust distribution \textcolor{MidnightBlue}{\citep{Zhao+2018,Koga+2019,Marchand+2020}}. 
For a distribution with a large number of small grains ($\lesssim 10$ $\mu$m), the resistivities become smaller as the grains adsorb charged particles. 
Early studies by \textcolor{MidnightBlue}{\cite{MellonLi2009,Li+2011}} assumed a distribution in which such small grains were present.
However, grains are expected to grow efficiently beyond $10$ $\mu$m in dense cores, which implies that higher resistivities are expected.
\textrm{Considering dust growth during the collapse, \cite{Lebreuilly+2023dust} showed how the resistivities can drastically change. 
Models by \cite{Tsukamoto+2023dust-resistivity} also show the effect of dust growth on the resistivities, but they also find a convergence towards a more steady distribution once the grains have grown beyond $1$ mm in size, such that the resistivities for ambipolar diffusion can be described by density-dependent power-laws of $\eta_{\rm AD}\propto n^{-0.5}$ for densities $\rho \lesssim10^{-13}$ g cm$^{-3}$.}
\textrm{Studying} the relative importance on the resistivities of the size distribution with the cosmic-ray ionization rate, \textcolor{violet}{\cite{Kobayashi+2023}} find that the size distribution is \textrm{a less significant factor compared to the cosmic-ray ionization rate}.

\subsubsection{Incorporation of dust dynamics and growth in collapse models}
In the context of planet formation, essential ingredients to incorporate in the models are dust dynamics and dust growth.
This is particularly important considering the growing evidence for an early onset of planet formation.
Some (magneto-)hydrodynamical models started to include the dynamics of dust in young forming disks in spherical collapse simulations. 
\textrm{\textcolor{MidnightBlue}{\cite{Vorobyov+2019}} and \textcolor{violet}{\cite{Bate2022}}, independently} carried out hydrodynamical simulations of disk formation in which they included the dynamics of dust particles during disk formation. 
Consistent with analytical predictions, they find size-dependent radial drift of dust particles.
\textcolor{MidnightBlue}{\cite{Lebreuilly+2020}} demonstrated that dust drifts toward the inner part of the disks with larger grains accumulating in the inner parts of the disk, where they cause an enhanced dust-to-gas ratio.
\textrm{More recently, \textcolor{MidnightBlue}{\citep{Vorobyov+2024}} followed up on their earlier work that already included dust growth and drift, by also accounting for the back-reaction of the dust on the gas \citep{Stoyanovskaya+2018}.
They concluded that the dust-to-gas ratio becomes high enough for the onset of planetesimal formation after only $\sim20$ kyr.
Considering the different properties of dust particles depending on their size and the underlying gas density in this context, \cite{Stoyanovskaya+2020} highlighted the importance of the underlying Mach number on the assumption of the drag coefficients.}

\textcolor{violet}{\cite{Tsukamoto+2021}} also modeled the drift of dust particles during the collapse, but also considered grain growth.
In contrast to \textcolor{MidnightBlue}{\cite{Lebreuilly+2020}}, they find that a fraction of the dust particles can be elevated by an outflow in the inner part of the disk,
become entrained in the envelope, and eventually fall back onto the outer part of the disk.
This idea is conceptually similar to earlier scenarios suggested to explain the transport of the oldest solids in the solar system, namely CAIs and chondrules \citep{Shu+1996,Shu+1997}. 
\textcolor{violet}{\cite{Tsukamoto+2021}} envision a scenario of multiple cycles that contribute to grain growth.
The dust grains drift through the disk, grow during the drift phase, are ejected through an outflow in the inner disk and 
fall back onto the outer disk as a larger grain. 
Considering multiple cycles of this mechanism, this could lead to a grain size distribution in the disk that is shifted towards larger grains
and which should be imprinted in the dust opacity spectral index of the envelope around the forming star-disk system.
Considering the meteoritic record, the transport mechanism could potentially explain the imprints of reprocessing reported for some chondrules.
Recently, \cite{Cacciapuoti+2024} followed up on this transport scenario by measuring the dust opacity spectral index in a few cores with outflows. 
Their results do not reveal an unambiguous correlation of dust growth with outflow strength though
and future observations are required to further test the scenario.

\section{Beyond isolated spherical collapse}
While variations in disk sizes between different models with non-ideal MHD are often solely explained as a sign of the underlying non-ideal MHD resistivities, 
another aspect tends to be overlooked, namely the role of the initial and boundary conditions.
It is a challenge to draw conclusions about the importance of individual effects from results that were obtained using different model setups codes.
Some groups start from initial conditions assuming a density profile according to a Bonnor-Ebert (BE) sphere, whereas others start with a uniform density distribution.
Considering this general issue, the parameter study by \textcolor{MidnightBlue}{\cite{Machida+2014}} stands out in terms of constraining the role of various effects. 
They adopted several model setups that were previously used by groups \textcolor{MidnightBlue}{\citep{Li+2011,Krasnopolsky+2012,Machida+2011,HennebelleCiardi2009,Joos+2012,Seifried+2012}} and recomputed those models with their own code. 
Activating the same non-ideal MHD effects and starting from similar initial mass-to-flux ratios, they found significant differences in disk formation depending on the assumption of the initial density profile.
Models starting with a uniform density distribution led to disk sizes of the order of 10 au, while models starting with a density distribution of a Bonnor-Ebert sphere yield larger disk sizes of about 100 au.
They also emphasized the role of the accretion recipe onto a sink. 
Under identical initial conditions, allowing the sink to accrete from a smaller region in its vicinity favors the formation of larger disks compared to a sink that is allowed to accrete from a larger region.

These results point to a more fundamental issue of modeling individual star formation as the outcome of the collapse of an isolated sphere. 
In fact, \textcolor{MidnightBlue}{\cite{Larson1969}} stated already that they assumed ’the simplest assumptions’ on the boundary condition and they ’again adopted the simplest assumptions’ for the initial conditions in the spherical collapse scenario.
While the assumption of spherical collapse has proven to be very helpful in constraining the effect of various physical parameters during collapse, it is important to keep in mind that reality can differ significantly from this idealized approximation.
It seems that the assumption of spherical symmetry is a fair assumption for the earliest phase of protostellar collapse corresponding to the formation of the first and second core, and model predictions about the properties of protostars during the first few thousands of years are likely to be very accurate - except for the uncertainty of the ionization rate at these early stages.

However, (almost) all of the observed disks are significantly older than $10^3$ yr -- even those associated with protostars that are classified as Class 0 objects.   
Observations revealed that stars are embedded in filamentary molecular clouds, and that they, at most, rarely form in isolation. 
Their formation and evolution is influenced by the environment in which they form and evolve. 
Variations of the magnetic field strength, the level of turbulence or the cosmic-ray ionization are factors that determine disk formation.
These effects can be incorporated in spherical collapse models by varying the initial conditions within the core,
relying on the crucial assumption that once a core has formed with specific initial conditions it can be considered as being detached from the dynamics and processes in the molecular cloud.

We know, however, from observations that the morphology of prestellar cores is significantly affected by the underlying dynamics in filamentary Giant Molecular Clouds \citep{Andre+2014,Kainulainen+2017,Pineda+2023,Hacar+2023}.
This implies to investigate the process of disk formation with a different approach than the spherical core setup has become timely.
As of today, there are relatively few studies compared to the large number of classical collapse models that account for larger-scale dynamics, such as infall or binary interaction, in the context of disk formation.
Both stellar encounters as well as infall have been modeled in computing-intense zoom-in simulations as well as in simplified model setups that allow to carry out cheaper models in terms of computing-time. 
In the following, we will summarize the efforts made in recent years distinguishing between the progress made in models that were configured to model specific scenarios through parameter studies and compute-intense models that aim for resolving the processes in multi-scale simulations.

\subsection{The role of infall}
Considering the presence of accretion streamers even around presumably evolved stars,
it is becoming increasingly evident that disk formation models need to take into account for the possibility of infall that is often anisotropic \textcolor{MidnightBlue}{\citep{Kuffmeier+2017,Kuznetsova+2019,Kuznetsova+2020}}.
Currently most of the numerical constraints on infall are from models that model the cloud dynamics, but do not resolve the formation of disks. 
\textcolor{MidnightBlue}{\cite{Padoan+2014}} demonstrated that infall through Bondi-Hoyle accretion can induce accretion bursts that are a prominent explanation for the luminosity problem.
Moreover, in line with the interpretation of the inertial flow model proposed by \cite{Padoan+2020},
\textcolor{MidnightBlue}{\cite{Pelkonen+2021}} showed that a significant amount of the material accreting onto the star was initially not gravitationally bound to the core,
and the relative mass fraction of accreting material scales with the final mass of the star.
This is also consistent with earlier results by \textcolor{MidnightBlue}{\cite{Smith+2011}},
who reported a prolonged accretion history of the more massive stars in their hydrodynamical simulations of a turbulent cloud.
Following up on these results, \textcolor{MidnightBlue}{\cite{Kuffmeier+2023}} also showed that a Class II young stellar object can return to Class I or even Class 0 phase in the event of massive infall.
It remains to be self-consistently modeled how the disk reacts to such events. 
However, there are parameter studies that considered the effect of individual infall events on the properties of star-disk systems.

Starting from spherical collapse models, several groups considered the case of infall from the envelope onto an existing disk \textcolor{MidnightBlue}{\citep{Vorobyov+2015,Lesur+2015,Bae+2015,Kuznetsova+2022}}.
They all find that the infall can trigger instabilities in the disk.
Carrying out hydrodynamical simulations, \textcolor{MidnightBlue}{\cite{Vorobyov+2015}} showed that infall can trigger gravitational instabilities that trigger accretion bursts of the star.
\textcolor{MidnightBlue}{\cite{Bae+2015}} and \textcolor{MidnightBlue}{\cite{Kuznetsova+2022}} find that these infall events trigger Rossby-wave instabilities,
which \textcolor{MidnightBlue}{\cite{Kuznetsova+2022}} attribute as possible seeds for gap, ring and structure formation in disks. 
Considering the possibility of misaligned infall from the envelope,
\textcolor{violet}{\cite{Thies+2011}} demonstrated in hydrodynamical simulations of a collapsing sphere with differences in the angular momentum orientation of the outer infalling radial layer
that such infall can induce the formation of an outer disk that is misaligned with respect to the primordial inner disk.
Considering the indications for late, post-collapse infall,
\textcolor{MidnightBlue}{\cite{Dullemond+2019}} developed a model of cloudlet capture, in which a low-mass gaseous cloudlet encounters a star modeled as a point mass with an impact parameter.
As a result of the encounter the hydrodynamical simulations carried out with the \pluto\ code showed the formation of extended arm features
similar to structures seen around Herbig stars such as for instance AB Aurigae.
The formation of such an extended arm (streamer) was also reported in the cloudlet capture models by \textcolor{MidnightBlue}{\cite{Hanawa+2022,Hanawa+2024}}.
\cite{Kuffmeier+2020} followed up on the cloudlet capture scenario by carrying out simulations with the \arepo\ code,
in which they showed that such encounters cannot only lead to the formation of extended arms, but also to the formation of a second-generation disk.
In the presence of an already (or still) existing primordial disk,
such an event can lead to the formation of a system consisting of misaligned inner and outer disk \citep{Kuffmeier+2021},
which is observable as shadows in the outer disk in scattered light observations \citep{Krieger+2024}. 
SU Aur is the most prominent candidate that might undergo such an event today \citep{Ginski+2021}.
The cloudlet capture scenario has also been adopted by \textcolor{MidnightBlue}{\cite{Unno+2022}}, who considered an encounter of a magnetized cloudlet.
They showed that the event can lead to magnetic acceleration of the existing inner disk for a favorable orientation of the magnetic field of the cloudlet compared to the field orientation in the disk.

\subsection{Clump to disk}
\textcolor{violet}{\cite{Bate2018}} carried out 3D hydrodynamical simulations with $3.5\times 10^7$ particles to produce the first disk population synthesis study.
As an initial condition of the simulation, they assumed a sphere with radius 0.404 pc.
The mass of the cloud was set to 500 M$_{\odot}$,
which corresponds to a density of $1.2 \times 10^{-19}$ g cm$^{-3}$.
Simulating the extended range of scales came at the cost of reduced physics compared to state-of-the-art simulations of spherical collapse simulations.
Radiative transfer was included, but magnetic fields were not taking into account such that magnetic braking is not accounted.
The cloud was initialized with a supersonic velocity field to account for turbulence in the cloud by generating a divergence-free random Gaussian velocity field with a power spectrum scaling with wave number $k$ as $P(k) \propto k^{-4}$, which is the same recipe as used in previous works \textrm{by} \textcolor{MidnightBlue}{\cite{Ostriker+2001}} \textrm{and} \textcolor{violet}{\cite{Bate+2003}}.
The resulting mean Mach number was $\textsf{M}=13.7$.
The simulation was run for 1.2 free-fall times, which corresponds to about 220 kyr.
By the end of the simulations, it showed a diverse distribution of more than 100 disks.
Interestingly, there sample shows a non-negligible distribution of more exotic configurations such as misaligned or even counter-rotating disks that are the result of infalling material after the initial formation phase or form via interaction with other stars through stellar encounters.
Moreover, the cloud dynamics led to the formation of circumbinary and circummultiple disks.
Using the same setup as \textcolor{violet}{\cite{Bate2019}}, \textcolor{violet}{\cite{ElsenderBate2021}} investigated the role of the metallicity on the disk distribution.
They used metallicities of 0.01, 0.1, 1, and 3 times the solar metallicity, and they find that disk sizes decrease for decreasing metallicities in the clump. 
In a further follow-up study, \textcolor{violet}{\cite{Elsender+2023}} investigated the frequency of circumbinary disks and they find that $90 \%$ of close binaries (separation less than 1 au) host disks.
Their distribution of circumbinary disks is bimodal with a second enhanced fraction of about 50 $\%$ for binaries with a separation of about 50 au.

In the case of \textcolor{violet}{\cite{Bate2018}}, simulating the extended range of scales came at the cost of reduced physics compared to state-of-the-art simulations of spherical collapse simulations. 
Radiative transfer was included, but magnetic fields were not taking into account such that magnetic braking is not accounted. 
\textcolor{MidnightBlue}{\cite{Lebreuilly+2021}} followed a similar approach of modeling the formation of stars dynamics of a massive turbulent sphere with an initial mass of 1000 M$_{\odot}$ and radius of $R_0\sim 0.38$ pc, but with magnetic fields included. 
The cloud is initialized with a ratio of thermal-to-gravitational energy of $\alpha = \frac{5 R_0 k_{\rm B} T_0}{2 GM_0 \mu_{\rm g} m_{\rm H}}$ with Boltzmann constant $k_{\rm B}$, a defined temperature $T_0$, clump mass $M_0$, mean molecular weight $\mu_{\rm g}$ and the mass of a hydrogen atom $m_{\rm H}$.
The simulations include magnetic fields and they account for ambipolar diffusion using the adaptive mesh-refinement MHD code \ramses. 
The initial average Mach number is set 7 and the mass-to-flux ratio is set to $\mu=10$.
Overall, \textcolor{MidnightBlue}{\cite{Lebreuilly+2021}} draw a similar conclusion to \textcolor{violet}{\cite{Wurster+2019nocatastrophe}} that disk formation is a frequent outcome of star formation regardless of ideal MHD or non-ideal MHD. 
Interestingly, there is no significant difference in mean disk size between the ideal and non-ideal MHD run, the mean disk size for the ideal MHD runs are even slightly larger. 

The authors caution nonetheless state that non-ideal MHD is important for disk formation. 
Based on the results of collapse simulations with higher resolution \textcolor{MidnightBlue}{\citep{Joos+2013}} they expect that magnetic braking will be more efficient and the disks therefore smaller if the resolution at the highest levels was higher than $\sim1$ au. 
In a follow-up study, \textcolor{MidnightBlue}{\cite{Lebreuilly+2024}} extended the parameter space by carrying out 5 additional simulations with similar setups. 
In one simulation, they carried out the identical simulation, but increased the accretion luminosity efficiency of sink particles from 0.1 to 0.5. 
The result is an increase of radiative feedback, which leads to more thermal support and less fragmentation in the clump. 
The resulting number of sink particles in the run with increased accretion luminosity efficiency is very similar to the result obtained for the ideal MHD run, but accretion luminosity efficiency of 0.1. 
As expected, an increase of the initial mass-to-flux ratio from 10 to 50 results in an increase of the final number of stars by more than 50 \%.
The star-formation efficiency increases by more than a factor 2 for runs with half the initial mass (500 M$_{\odot}$ instead of 1000 M$_{\odot}$), and hence twice the initial thermal-to-gravitational energy ratio.
Finally, the authors report an increase in star-formation by about 40 \%, when adopting a barotropic equation-of-state instead of treating the radiative transfer with the flux-limited-diffusion recipe \textcolor{MidnightBlue}{\citep{Commercon+2011,Commercon+2014}}.
Regarding the disk population, the authors conclude that disk formation is ubiquitous for all setups. 
In addition, the disk population is practically the same regardless of account for outflows or not although outflows reduce the accretion rate, and hence the luminosity of the protostar \textcolor{MidnightBlue}{\citep{Lebreuilly+2024outflows}}.
Similarly to the hydrodynamical models by \textcolor{violet}{\cite{Bate2018}} the derived disk sizes are in approximate agreement with observations regardless of ideal or non-ideal MHD \citep[e.g.,][]{Maury+2019,Sheehan+2022}, while the disk masses are systematically higher in the models compared to disk masses derived from observations \citep{Sheehan+2022}. 

Comparing the ambipolar diffusion runs with mass-to-flux ratios of $\mu=10$ and $\mu=50$, the study shows that the disk sizes in the less magnetized run are significantly larger by about $30 \%$ to $50 \%$, which confirms \textrm{the effect of magnetic fields on the disk size during their formation.} 
Considering the difference between the models with ideal MHD and ambipolar diffusion, the authors point out, however, that the conditions in the disks are significantly different. 
In the ideal MHD case, the magnetic field strength in the disk is significantly higher implying plasma-beta values of $\beta<1$ in the disk, while the disks in the ambipolar diffusion case are generally dominated by thermal pressure with $1<\beta<100$.

\subsection{Giant Molecular Cloud to disk}
While the clump-models \textrm{by} \textcolor{violet}{\cite{Bate2018,ElsenderBate2021}}
\textrm{as well as} \textcolor{MidnightBlue}{\cite{Lebreuilly+2021,Lebreuilly+2024}} allowed to derive disk population synthesis, the models starting from an initial spherical clump of $<0.5$ pc cannot truly account for the Giant Molecular Cloud dynamics. 
While it is common practice to study star formation in molecular cloud simulations, almost all models do not have the resolution to resolve the scales necessary to form disks. 
The zoom-in simulations by \textcolor{MidnightBlue}{\cite{Kuffmeier+2017}} mark an exception in this regard. 
They modeled the dynamics of a Giant Molecular Cloud including supernova feedback from massive stars that drives the turbulence in 3D simulations of a cubical box with 40 pc in length 
and assuming periodic boundary conditions. 
The simulations included magnetic fields and solved the equations of ideal MHD. 
Using the zoom-in technique, they then resolved the formation of 6 stars with highest resolution of 2 au, 
and of 3 additional stars with highest resolution of 8 au.
The temperature in the cloud is computed using tables for heating and cooling, 
though the gas at highest densities corresponding to the inner $\sim100$ au is treated as quasi-isothermal with a temperature of 10 K. 
The study showed a diversity of the disk formation process and demonstrated that star formation is a heterogeneous process depending on the parental environment. 
In fact, the models predicted the frequent occurrence of filamentary accretion channels that feed the young disk with fresh material from the environment.
On scales of 10 to $\sim100$ au, this filamentary mode of accretion through streamers is consistent with results from previous protostellar collapse models with turbulence as first prominently reported by \textcolor{MidnightBlue}{\cite{Seifried+2013}}
and as also seen by other groups \textcolor{MidnightBlue}{\citep{Offner+2010,Joos+2013,Li+2014,Heigl+2024}}. 
On larger scales, the star-disk systems forming in more turbulent and massive environments are also fed with additional material through filamentary arms on core scales and beyond $\sim 1000$ au to $10000$ au.
Independently, \textcolor{MidnightBlue}{\cite{HeRicotti2023}} carried out zoom-in simulation for massive cores of $\sim10$ M$_{\odot}$ to $\sim100$ M$_{\odot}$ in size starting from a Giant Molecular Cloud down to a disk forming around a massive star using radiative transfer MHD simulations.
The dynamical range is similar in the high resolution case ($\approx 7$ au) and the study also reveals a connection of the forming disks to the molecular cloud environment through filamentary arms of $\sim1000$ to $\sim10000$ au in length.
Considering bridge structures of $\sim 1000$ au in length similar to those observed for IRAS16293, \textcolor{MidnightBlue}{\cite{Kuffmeier+2019}} showed and \textcolor{MidnightBlue}{\cite{Lee+2019}} \textrm{independently} confirmed that such structures can occur as transient phenomena during protostellar multiple formation in a turbulent environment.

\begin{figure}[h!]
\begin{center}
\includegraphics[width=12cm]{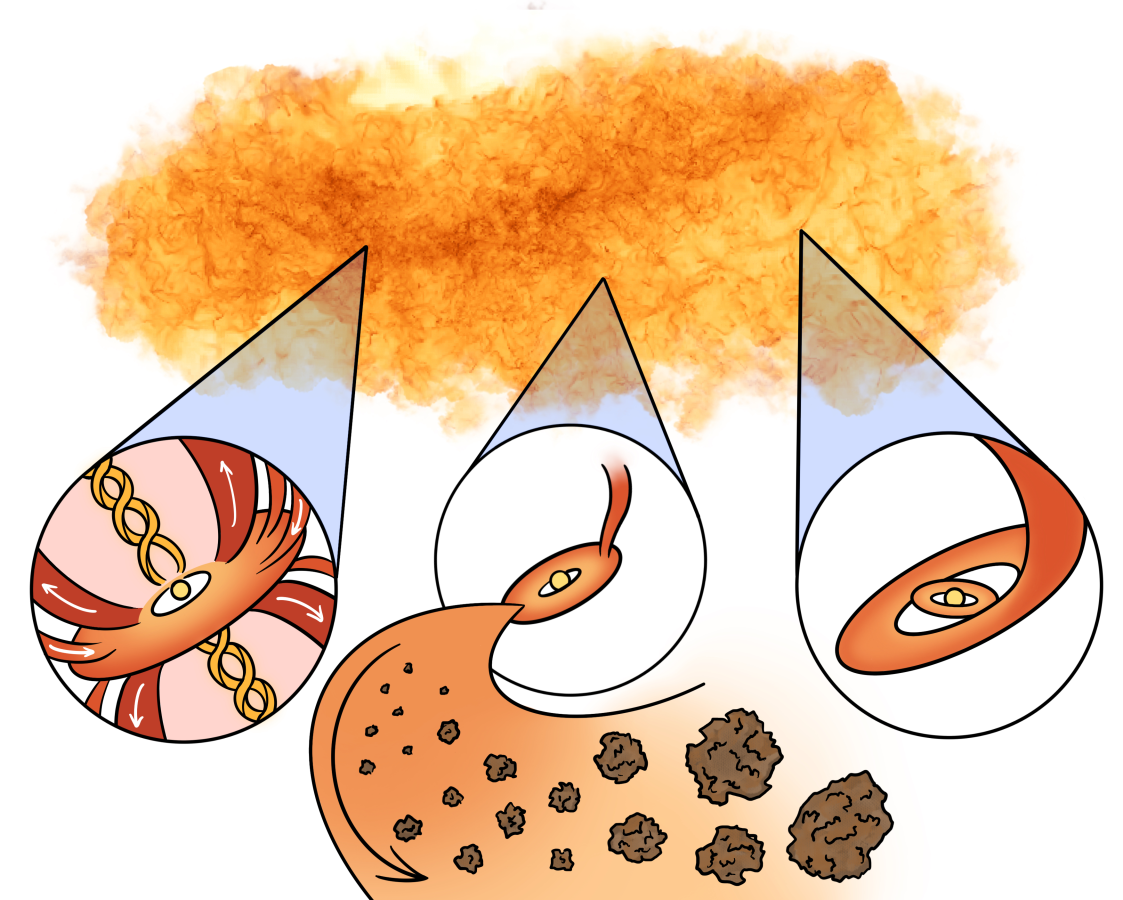}
\end{center}
\caption{Illustration of heterogeneous star formation and possible outcomes of disk formation in a Giant Molecular Cloud. 
The upper panel shows the density fluctuations in a filamentary cloud based on data from \textcolor{MidnightBlue}{\citep{Kuffmeier+2017}}. 
The remaining part sketches different stages of the disk.
The left inset shows a young disk with a strong magnetically-induced outflow \textrm{enclosing} a jet, the middle inset shows infall through a streamer solely delivering material, while the right inset shows the scenario where infall leads to the formation of a misaligned outer disk during a late infall event. 
The cartoons were made by Martine Lützen, a former Master student at the Niels Bohr Institute.
}
\label{fig:LutzenKuffmeier}
\end{figure}

\textcolor{MidnightBlue}{\cite{Kuffmeier+2018}} also carried out a zoom-in simulations for one of the stars with a resolution as high as 0.06 au for a time interval of 1000 years at about 50 kyr after star formation. 
These simulations resolved the infall of a gas blob onto the disk that triggered a gravitational instability that was responsible for an accretion burst. 
While the dynamical range of the simulation allowed to take into account for the larger scale environment, 
important physical effects such as non-ideal MHD and/or a more realistic treatment of the thermodynamics was lacking in these models compared to state-of-the-art models of spherical collapse. 
The challenge for the upcoming decade will be to fill the gap between spherical collapse models with advanced multi-physics treatment, 
but highly idealized initial and boundary conditions, 
and multi-scale models that self-consistently account for the Giant Molecular Cloud dynamics, but lack advanced treatment of relevant physical processes.
The improvements of infrastructure of more powerful supercomputers together with optimized numerical codes that enable more efficient computing \citep[e.g.,][]{Hopkins2017,Nordlund+2018,Price+2018,Stone+2020,Weinberger+2020} allow us to accept this challenge.
\textrm{This also includes the possibility to account for stellar feedback mechanisms that can shape and affect disk formation. 
For instance, UV radiation feedback plays an important role for the overall cloud dispersal as shown in radiation hydrodynamics models \textcolor{MidnightBlue}{\citep{Kim+2018, Fukushima+2020}}.  
Together with implementations of protostellar outflows, e.g., \cite{Guszejnov+2021} or \textcolor{MidnightBlue}{\cite{Lebreuilly+2024outflows}}, these are potentially important effects shaping the accretion process of stars and thereby their disks.
More recently, as part of the STARFORGE initiative \citep{Grudic+2021Starforge}, \cite{Guszejnov+2022} already demonstrated the possibility to study supernova feedback, stellar radiation, protostellar jets and winds together in one simulation. 
Continuing these efforts while applying high enough resolution} will allow us to test the frequency and properties of various outcomes of infall-induced features such as misaligned disks or disk instabilities that are possible outcomes of a heterogeneous star formation process happening in Giant Molecular Clouds (see illustration in \Fig{LutzenKuffmeier}).

\section{Reflections and outlook}
Since the first models of protostellar collapse of a spherical core \textcolor{MidnightBlue}{\citep{Larson1969}},
there have been a lot of successful efforts in improving our understanding of protostellar formation. 
It is possible to resolve the collapse with a resolution that even allows to model the dynamics within the inner 1 au of the forming protostar,
while there also as has been systematical incorporation and testing of additional physical effects. 
Starting from the crucial assumption for disk formation of initial rotation, models have improved to a degree where it is possible to account
for more subtle effects such as the role of the ionization rate or the dust size distribution
on the resistivities, which modify the efficiency of magnetic braking of the forming disk.

Regarding initial and boundary conditions, the assumption of spherical collapse of an isolated prestellar core has proven to be successful as a the commonly used, fiducial assumption to test and compare the parameter space.   
In fact, the early ALMA images of dust continuum \citep{ALMAPartnership2015,Andrews+2018}, despite their substructures and diversity, 
gave the impression as if the disks are indeed isolated entities that can be considered detached from the environment in which they are embedded in.
However, today we know that there is a severe observational bias in observing dust emission at relatively large wavelength of about 1 mm. 
\textit{Gas} observations of disks, reveal a different picture \citep{Oeberg+2021} than dust continuum images obtained at $\sim1$ mm wavelength. 
First, the gaseous parts of disks are significantly larger and less structured.
The smaller size of the dusty parts of disks can be explained by radial drift of dust particles \citep{Weidenschilling1977}.  
Second, and of fundamental importance for our assumptions on the initial and boundary conditions of our models, there is clear evidence of filamentary arms, nowadays commonly referred to as streamers \citep{Pineda+2023}. 
It is especially striking that HL Tau, the prime target of ALMA, hosts a prominent streamer \citep{Yen+2019} that impacts the disk \citep{Garufi+2022}. 
By now, observations of various chemical tracers confirmed streamers on multiple scales ranging from small spiral-like features of a few 10 to 100 au in length around young disks with ongoing dust growth (Oph IRS63) \citep{Segura-Cox+2020,Segura-Cox+2023,Flores+2023} to arcs that are several 1000 au in length associated with prestellar cores (Per-emb-2) \citep{Pineda+2020}.
There is also observational evidence for streamers associated with ongoing simultaneous star and planet formation in the cases of [BHB2007]-1 \citep{Alves+2020}, and (Per-emb-50) \citep{Valdivia-Mena+2022}. 
\textrm{A systematic search of streamer candidates in NGC 1333 revealed a fraction of $\approx40 \%$ associated with young protostellar sources \citep{Valdivia-Mena+2024}.}
The detection of large-scale streamers that are supposed to replenish the mass reservoir for star and disk formation is consistent with modeling results that predict infall of a substantial fraction of the accreting mass that was initially not bound to the prestellar core \textcolor{MidnightBlue}{\citep{Smith+2011,Pelkonen+2021,Kuffmeier+2023}}.
In addition, models of star formation in a Giant Molecular Cloud show that for many stars a significant amount of the final stellar mass accretes late (i.e., $>$100 000 years after stellar birth) and it is considered as a probable solution to the luminosity problem of protostars \textcolor{MidnightBlue}{\citep{Padoan+2014}}.
It is consistent with observations of streamers around more evolved Class II Young Stellar Objects such as AB Aur \citep{Grady+1999}, SU Aur \citep{Akiyama+2019,Ginski+2021}, GM Aur, \citep{Huang+2021}, Elias 2-27 \citep{Paneque-Carreno+2021}, DR Tau \citep{Mesa+2022,Huang+2023}, RU Lup \citep{Huang+2020}, DO Tau \citep{Huang+2022} as well as strong indications of late infall from statistics of reflection nebulae \citep{Gupta+2023}.

\textrm{As elaborated in section 3.1, the late addition of gas with substantial angular momentum even offers an additional path to disk formation beyond the initial protostellar collapse phase. 
In agreement with earlier suggestions by \textcolor{MidnightBlue}{\cite{Padoan+2005}}, \textcolor{MidnightBlue}{\cite{Kuffmeier+2023}} demonstrated that the larger contribution of late infall can explain the subtle trend of larger disk sizes for increasing mass of the corresponding host stars seen in CO observations \citep{Long+2022}. 
Furthermore, the infall of material is typically misaligned with respect to the orientation of the star-disk system \textcolor{MidnightBlue}{\citep{Kuffmeier+2024}}, especially at later stages \textcolor{MidnightBlue}{\citep{Pelkonen+2024}}, leading to a more chaotic pathway of star-disk formation \textcolor{violet}{\citep{Bate+2010}}.
Most recently, conceptual papers followed up on the possibility of post-collapse disk formation arguing that a significant amount of observed Class II disks are in fact the result of prolonged Bondi-Hoyle like accretion in the turbulent interstellar medium \citep{Padoan+2024,Winter+2024}.
If this mode of second-generation disk formation proves to be significant, it also implies that interpreting surveys of stellar age-dependent disk fractions \citep{Haisch+2001,Mamajek2009,Richert+2018} solely as an outcome of the evolution of either viscous or wind-driven disks \citep[e.g., review by][and references therein]{Manara+2023} is misleading. 
In the post-collapse infall picture, the lower fraction of disks around more evolved stars instead reflects the decreasing probability of experiencing mass replenishment through late gas encounters.
Disks around older stars should instead be considered as either disks that experience prolonged mass replenishment or as second-generation disks that formed several million years after the initial protostellar collapse phase through post-collapse gas encounters.}

Considering the compelling observational evidence, we have recently started to consider filamentary accretion via streamers as part of the disk formation process. 
The effect of infall on the processes in the disk and their impact on planet formation has only been poorly investigated yet, though infall through streamers may be of key importance \textcolor{MidnightBlue}{\citep{Bae+2015,Lesur+2015,Kuffmeier+2018,Kuznetsova+2022}}.
In particular, infall likely plays an important role in: 
\vspace{-0.03cm}
\begin{itemize}
   \item regulating the disk size, 
   \item triggering instabilities in young disks and thereby initiating substructures, 
   \item inducing misaligned disks visible as shadows in scattered light observations,
   \item seeding finite amplitude pressure traps,
   \item potentially modifying the chemical composition of planetary systems,  
   \item resetting the disk entirely.
\end{itemize}
\vspace{-0.03cm}
Also we commonly use the term `streamer' regardless whether it describes filamentary infall on scales of $\sim$10 au onto a very young disk or a $\sim10^3$ au elongated arm associated with an evolved YSO Class II object. 
All streamers, to some extent, reflect the turbulent nature of Giant Molecular Clouds, but there are indications from modeling that there are differences in their origin. 
While small-scale accretion streamers around young stars were already seen in spherical collapse models of individual protostars, larger scale arms around more evolved stars were best described by a star capturing material during an encounter with dense gas, for simplicity assumed as a cloudlet of gas in early models \citep{Dullemond+2019,KuffmeierGoicovicDullemond+2020}.
This suggests that the small-scale streamers might be a direct outcome of gravitational collapse with perturbations induced by turbulence, whereas the latter is analogous to Bondi-Hoyle like accretion with an impact parameter.
This interpretation would be consistent with the scenario that star formation is a two-stage process \textcolor{MidnightBlue}{\citep{Kuffmeier+2023}}, or three-phase when including the process leading to core formation in the sequence \citep[][]{Padoan+2020}, consisting of an early collapse phase followed by an optional post-collapse phase of material that is initially not gravitationally bound to the collapsing core. 
Apart from that, the majority of stars is associated with stellar multiples \citep{DuquennoyMayor1991,Connelley+2008,Raghavan+2010}, and multiplicity is already common among young, deeply embedded protostars \citep{Chen+2013,Tobin+2015}. 
In this context, it will be important to consider the possible effects and perturbations through stellar interactions that may well affect the properties of disks, too \citep[see for instance reviews by][and references therein]{Offner+2023,Cuello+2023}.

What is lacking at the current stage is a model that allows us to distinguish between the relevance and occurrence of the different modes without an intrinsic bias in the model setup that favors or even excludes possibly important scenarios.
A self-consistent model of the accretion process of star-disk systems with high resolution together with the molecular cloud dynamics for long enough times (i.e., beyond $\sim 1$ Myr) would fill this gap.
\textrm{Developing such models is particularly timely considering the progress in the field of planet formation.}
While it has been broad consensus that planets form in protoplanetary disks that themselves are a byproduct of the star formation process happening in Giant Molecular Clouds,
the traditional understanding of considering disk formation isolated from the molecular cloud dynamics still is the leading paradigm when considering planet formation.
This picture has been and is increasingly disrupted \textrm{in} recent years, such that we are moving to a more dynamic interconnected picture of star and planet formation. 
Since the advent of Herschel, it is clear that molecular clouds are filamentary in nature with cores generally forming associated with these filaments.
This implies that the individual cores are less isolated than in the traditional illustration of a protostar that forms at the center of a symmetrical core. 
Moreover, observations of rings and gaps in dust continuum images of disks are commonly interpreted as signs of ongoing dust growth, which is a necessary precursor of planet formation. 
This led to a paradigm shift towards \textrm{an} onset of planet formation \textrm{in embedded disks} as it is reflected by a change in terminology from referring to circumstellar disks as `planet-forming' rather than `protoplanetary'.
Accepting the revised picture implies that planet formation already \textrm{occurs} at a time when the disk is still in its formation phase and prone to infall. 
Nowadays, star and planet formation are therefore considered as processes that are much more interlinked than we used to think for many decades. 

From a technical point of view, the challenge for the upcoming years will be to connect state-of-the-art multi-physics models
with multi-scale models that currently lack parts of the physics that is important for disk formation on smaller scales. 
Future models will allow us to fill the currently existing gap in understanding to what extent the properties of disks are governed by the protostellar environment
and how frequent various outcomes such as infall-induced misaligned disks and instabilities are.
The prospects in succeeding with this task are bright. 
Some models have already succeeded in resolving young disks with a resolution of less than 0.1 au in the context of a magnetized Giant Molecular Cloud that is $\sim 10$ au in scale. 
Others provided first synthetic disk populations based on disk formation in a star-forming clump with non-ideal MHD. 
The increasing necessity to account for the star formation process against the background of a presumably early onset of planet formation in embedded disks 
together with the tremendous development of computational resources and techniques for the exascale era of supercomputing,
will presumably lead to rapid progress in our understanding of disk formation and evolution by the end of the decade.

\section*{Conflict of Interest Statement}
The author declares that the research was conducted in the absence of any commercial or financial relationships that could be construed as a potential conflict of interest.

\section*{Author Contributions}
MK wrote the article.

\section*{Funding}
The research of MK is supported by a H2020 Marie Sk\l{}odowska-Curie Actions grant (897524) and a Carlsberg Reintegration Fellowship (CF22-1014). 

\section*{Acknowledgments}
MK thanks the two referees for constructive feedback and useful suggestions that helped to improve the quality of the manuscript. MK also thanks the organizers of the workshop \textit{Simulating Physics in Celestial Ecosystems (SPiCE)} in Sendai, Japan, for an inspiring week full of exciting science and discussions. 

\section*{Data Availability Statement}
No new data is presented in this article.

\bibliographystyle{Frontiers-Harvard} %  Many Frontiers journals use the Harvard referencing system (Author-date), to find the style and resources for the journal you are submitting to: https://zendesk.frontiersin.org/hc/en-us/articles/360017860337-Frontiers-Reference-Styles-by-Journal. For Humanities and Social Sciences articles please include page numbers in the in-text citations 
%\bibliographystyle{Frontiers-Vancouver} % Many Frontiers journals use the numbered referencing system, to find the style and resources for the journal you are submitting to: https://zendesk.frontiersin.org/hc/en-us/articles/360017860337-Frontiers-Reference-Styles-by-Journal

%\bibliography{general.bib}

\end{document}